\documentclass[12pt]{article}

\usepackage{amsmath}
\usepackage{amssymb}
\usepackage{eufrak}

\begin{document}

\title{Theory of shell structure and of the ``magic" effect in spherical nuclei
\footnote{Zh. Eksp. Teor. Fiz. {\bf 61}, 1303--1318 (1971) [Sov. Pys. JETP
{\bf 34,} No.~4, 694-701 (1972)]}}

\author{V.G. Nosov$^{\dagger}$ and A.M. Kamchatnov$^{\ddagger}$\\
$^{\dagger}${\small\it Russian Research Center Kurchatov Institute, pl. Kurchatova 1, Moscow, 123182 Russia}\\
$^{\ddagger}${\small\it Institute of Spectroscopy, Russian Academy of Sciences, Troitsk, Moscow Region,
142190 Russia}
}

\maketitle

\begin{abstract}
A consistent theory is developed of the volume energy oscillations of spherical nuclei
due to sharpness of the Fermi distribution boundary for quasiparticles. The lowest
value of the oscillating part of the energy corresponds to a magic nucleus. A formula
is obtained for the corresponding limiting momentum of a quasiparticle and it is shown
that we have here an isolated point of a temperatureless second-order phase transition.
An expression for the discontinuity of the derivative of the energy of the body with
respect to the number of particles is obtained in the case of a sharp (step-like) Fermi
distribution limit. Comparison with experimental nuclear-mass data permits some
conclusions to be drawn regarding the true structure of the boundary layer of the Fermi
distribution and regarding its variation with increasing nuclear size. In the region of
magic nuclei actually accessible up to the present time, apparently no signs are observed
of any appreciably expressed residual phenomenon, such as the Cooper phenomenon, which
would result in instability of the energy spectrum of infinite nuclear matter with an
absolutely sharp Fermi limit for quasiparticles.
\end{abstract}

\section{Introduction. Macroscopic treatment of magic nuclei}

The study of the nucleus considered as a macroscopic body originated with the well known
work of von Weizs\"acker [1].  Recognition of the distinction between volume and
surface contributions to the binding energy was an important step in understanding the
saturation properties of nuclear forces and served as the basis for current ideas regarding
uniform nuclear matter (see, for example, Landau and Lifshitz [2]). The subsequent
inclusion of the electrostatic energy and addition of the symmetry term taking into account
the difference in the numbers of protons and neutrons [1] gave these ideas a rather
high degree of refinement which permitted the total binding energies of a very large number
of nuclei to be described with high relative accuracy. This, however, only deepened interest
in possible deviations from the Weizs\"acker formula, and the question of these deviations
has arisen repeatedly in one form or another during the past thirty-five years.  The most
easily perceived cause is the quantum-mechanical discreteness of elementary particles, i.e.,
simply the obvious fact that protons and neutrons can be added to a nucleus only in integral
numbers. This, naturally has stimulated the consideration of other similar deviations as
having a more or less clearly expressed quantum nature.

However, the question of which deviations from the Weizs\"acker formula have a
macroscopic origin lies essentially in a different plane. At the present time we have no
serious basis for rejecting such a possibility. On the opposite, developments in recent
years indicate that macroscopic quantum phenomena are rather typical for nuclei
which are not too light. In the first place, the simplest model of a uniform body of this
type with limited size---the independent-particle model with a constant general potential in
the internal region---permits one to make rigorous calculation of several important characteristics.
The instability demonstrated by Nosov [3] for a spherical configuration of this
model has an explicitly macroscopic nature. It is true that determination of the true stable
equilibrium shape of the body encounters serious difficulty that the problem has no small
parameter.  The attempts to ignore this fact, with resort to direct numerical calculations,
are difficult to analyze and we will not set for ourselves this goal. However, as often
happens in similar situations, experimental data come to our assistance. The ``collective"
macroscopic order of magnitude of quadrupole moments of non-spherical nuclei is well known;
an individual quasiparticle could not produce a quadrupole moment of this order.  Closely
connected with this fact is the usually smooth behavior of the deformation, i.e., the fact
that this macroscopic characteristic of an non-spherical nucleus is practically independent of
the individual quantum state of the last nucleon (quasiparticle) of the Fermi distribution.

Also of great significance is the study, intensely carried out during the last fifteen years,
of the rotation of the nucleus as a whole, and also of oscillations of the nuclear deformation.
Satisfaction of the laws of quantum mechanics does not diminish, of course, the macroscopic
nature of this type of motion. It is remarkable that, even from calculations based on a
simple model [3],  it can be clearly seen how the characteristics of the
corresponding collective degree of freedom (for example, rigidity of the spherical
configuration with respect to deformation) ``develop" with increase of the number of particles
in the nucleus.  The macroscopic properties of a body should lose their meaning and
physical content in the transition to too small number of particles.

The undoubted existence of spherical nuclei is not compatible with the primitive model
of ref.~[3], and the real situation in a nucleus must be more complex. The most important
reason for that is probably the residual interaction between the quasiparticles, which is
unavoidable in a body of finite size. The loss of equilibrium non-sphericity is difficult to
imagine without some qualitative change in the nuclear structure. However, as was shown by the
theory of the corresponding phase transition developed by Nosov [4], this complication of
the physical picture only increases the role of macroscopic quantum effects. Actually, our
attention is primarily attracted not only by the singularities of thermodynamic quantities
(as occurs for ordinary temperature-dependent second-order phase transitions but also by the
singularities of the macroscopic-body ground-state wave function (which depends
parametrically on the deformation), observed at the Curie point (see, for example, Eq.~(28)
in ref.~[4])\footnote{For ordinary condensed matter we can also give a physical
example of a temperatureless
second-order phase transition. When a single-domain ferromagnetic particle grows in size,
beginning at some critical size, a smooth change occurs in the spin directions, which
gradually destroys the uniformity of magnetization over the volume. The theory given by
Frei et al. [5] permits wave functions of the phases to be determined near a transition
point. It is easy to see that their properties are consistent with the semi-phenomenological
reasoning developed by Nosov [4] for the case of a nucleus.}.

It is possible that the significance of macroscopic quantum effects in a nucleus is still
somewhat underestimated. In any case, against the background of the progress described above,
the widely held opinion that the so-called magic numbers are clearly of non-macroscopic
origin [6] appears to be more a dogmatic assertion than a fact following directly from experiment.
It is natural to relate a theoretical study of the macroscopic properties of a medium with the
experimentally observed singularities of the thermodynamic variables. This facilitates
observation and detailed investigation of characteristic phenomena such as phase transitions.
A specific feature of a nucleus is the direct accessibility of absolute zero temperature
(the ground state), and the most important thermodynamic variable turns out to be the
corresponding energy, i.e., the nuclear mass. As far as can be judged from the experimental
data shown schematically in Fig.~1, the singularities under discussion are apparently breaks,
i.e., discontinuities in the derivative of the energy with respect to the number of particles.
Where the curve becomes sluggish, almost horizontal, on one side of the singularity, we are
entering the region of non-spherical nuclei, and the Curie point corresponds to the phase
transition discussed by Nosov [4], a consequence of which, in particular, is a
change in the equilibrium shape of the nucleus. In addition, in Fig.~1 we can clearly see
the magic cusps, i.e., downward spikes at whose vertices are located the corresponding magic
nuclei. We emphasize the difference of this singularity from the ordinary Curie point. A
qualitative difference between the phases---different values of the chemical potential---arises
only near an isolated transition point. Away from this point it is not possible to point to a
quality which would be possessed, say, by one of the phases while the other would not have it.
The actual state of the body apparently does not undergo a discontinuity (with macroscopic
accuracy, of course; see the following section), so that in other respects we appear to be
dealing with a second-order transition.

In opposition to the impression created by Fig.~1, the widely used explanation of magic
numbers [6] appeals directly to the pattern of filling by fermions of $(2j + 1)$-fold
degenerate single-particle levels in the field of some spherically symmetric potential.
After a shell is filled, the next nucleon approaches the lower edge of the region of
continuum states and the nucleon binding energy $\varepsilon$ (the chemical potential taken
with a reversed sign) correspondingly drops. The weak point of this type of interpretation
is that a similar situation should arise also after the filling of each level individually,
and not only of an entire definite group of levels (a shell). In comparison with the
gaps separating neighboring shells, the distances between the levels inside each shell
cannot always be small; there is no way in which such a small parameter
could be obtained in this case. Specific calculations of single-nucleon level schemes
confirm the validity of this reasoning (see, for example, the neutron level scheme from
the monograph by Hodgson [7]). As a rule, the gap between shells is only two or three
times greater than the typical distance between levels, and we would have to find a whole
series of closely spaced cusps corresponding to the filling of each of the levels. In
reality, for not too light nuclei, beyond the magic number 28, only the cusps shown in
Fig.~1 are observed, which correspond to the much less frequently encountered ``true" magic
numbers. It is symptomatic that even a small, far from qualitative, excess of the gaps
between shells over the level spacing inside the shells is achieved by selection of a
rather substantial number of parameters. The gaps following shells 2 and 8 turn out to be
large for practically any reasonable potential [6], but for the magic numbers 28, 50, 82,
and 126 it is already impossible to make this statement. The fact that even the average
field acting on the nucleon is determined here by no less than four parameters (the
potential-well radius and depth, the width of the transition layer at the surface,
and the intensity of the spin-orbit interaction within its boundaries) naturally provides
additional cause for skepticism. In final analysis, we can attempt to interpret almost
any phenomenon as the result of appropriate selection of fitting parameters; however, we have a
right to expect more from the theory.  Therefore, the widespread explanation of magic
numbers produces seems to be somewhat artificial\footnote{Here and subsequently
we will refrain from making any far reaching parallels with the
shell structure of an atom. The principal distinction from nuclear matter is due to the
Coulomb field of the nucleus at the center, as a result of which the atom has a sharply
expressed spatial non-uniformity.}.

The key to understanding the macroscopic nature of the magic cusps and the nature of the
phenomenon itself, which does not depend on specific numerical values of the nuclear
parameters (equilibrium density of nuclear matter, structure of the transition surface
layer, value of spin-orbit coupling, and the like), lies in a well-known general property
of Fermi systems.  The thermodynamic variables of these systems (for example, the total
magnetic moment, the energy, or number of particles $N$) usually reduce to sums over a
large number of fermion-filled states characterized by definite sets of quantum numbers.
The simple replacement of summation over quantum numbers by integration is a crude
treatment which leads to a result depending smoothly on such variables as the limiting
momentum $k_f$ of the Fermi distribution or the intensity of the applied magnetic field.
However, a rigorous calculation of the sum by means of Poisson's formula [8]
\begin{equation}\label{1}
    \sum_{n=0}^{\infty}\varphi(n)=\frac12\varphi(0)+\int_0^{\infty}\varphi(n)dn+
    \sum_{\nu=1}^{\infty}\int_0^{\infty}\left(e^{i2\pi\nu n}+e^{-i2\pi\nu n}\right)
    \varphi(n)dn
\end{equation}
shows that the thermodynamic variables, generally speaking, undergo oscillations as a
result of the third term in (\ref{1}). Here the main contribution is from the quasiparticles
closest to the Fermi surface. Actually, the oscillating integral which
occurs inside the summation over $\nu$ would be extremely small if $\varphi(n)$ were a
sufficiently smooth function. The latter, however, contains as a factor the statistical
distribution of quasiparticles, which changes sharply in the vicinity of the Fermi surface.
An increase in temperature smooths the $\varphi(n)$ function, as a result of which the
oscillating effect falls off exponentially. A well-known specific example of this
phenomenon is the de Haas-van Alphen effect [9], i.e., quantum oscillations of the
magnetic susceptibility of metals at low temperatures.

Since we have in mind first of all nuclear applications, it is particularly important to
emphasize that, in addition to the sharpness of the Fermi distribution, there is another
equally necessary condition. An oscillation clearly does not arise with only a single
summation according to Eq.~(\ref{1}), i.e., if $n$ is the only quantum number of the particles
being discussed (in the general case, of quasiparticles). An elementary calculation shows
that in the one-dimensional case the last term in (\ref{1}) describes only the aforementioned
trivial fact of addition of particles one at a time; physically this corresponds to the
absence of a macroscopic effect\footnote{A similar situation occurs also in a
three-dimensional potential well of sufficiently
irregular shape so that the particle energy is its only quantum number.}.
An important role is therefore played by the
fact that the quasiparticle in the spherical nuclei discussed below has an orbital angular
momentum $l$.  From the calculations of the next section it is evident that
there are enough quantum numbers in a spherical nucleus for occurrence of oscillations of
a macroscopic scale. However, upon formal application of relation (\ref{1}), there can arise
also terms which are not macroscopic in nature, and which must be discarded. In making this
distinction the decisive criterion is
\begin{equation}\label{2}
    \rho_f=k_fR\gg 1
\end{equation}
($R$ is the nuclear radius); for brevity we will also refer
to this dimensionless parameter as the limiting momentum of the Fermi distribution.

To avoid misunderstanding, we note that the uses of Eq.~(\ref{1}) in solid-state physics and
in nuclear physics have somewhat different aims. In the study of metals by means of the
de Haas—van Alphen effect, problems of sharpness of the Fermi surface essentially do not
arise, and the problem is to determine the shape of the Fermi surface in momentum space
(more accurately, in quasi-momentum space) [9]. On the other hand, in the spherical
nuclei of interest to us, the obvious isotropy removes the question of the ``shape" of the
Fermi limit, but its detailed structure is nowhere near as clear, owing to the residual
interaction (see above).  Furthermore, it is apparently only because of the residual
interaction that spherical nuclei in general exist---we noted above that in a model without
the interaction [3] a sphere is unstable. However, the effect of this circumstance on
the energy of the body as a whole and on the structure of its quantum state in the region
of rapid fall-off of the quasi-particle distribution function, and also the limiting behavior
in accordance with which the residual interaction may disappear with increasing nuclear radius,
represent a field which is still little studied.  Therefore, in the calculations which follow,
we will take tentatively the ordinary Fermi distribution function unperturbed by the residual
interaction. In the current state of the problem, only experiment can answer the question as
to how closely we approach the ideal limit of a step-shaped distribution for the nuclear sizes
achievable at the present day. Sections 3 and 4 of the present article are devoted to questions
bearing on this point.

For what follows it is very important to comment on the various characteristics of the
quasiparticle and the possible relation between them.  The point is that in comparison with
the primitive model of a Fermi gas in an external field, a real Fermi liquid (i.e., a bound
system of fermions of finite size) has some qualitative differences in this respect. We have
attempted to compare these situations schematically in Figs.~2a and 2b, respectively. In both
cases we can speak of an energy, if we reckon it from the zero of the kinetic energy of an
external free nucleon. At the limit of the Fermi distribution this quantity reduces to the
chemical potential $\varepsilon_f=-\varepsilon$, where $\varepsilon$ is the reversible work function
of the particle for removal from the nucleus (the nucleon binding energy). In addition, in
the case corresponding to Fig.~2a, there exists also an energy $\varepsilon'$ relative to the
bottom of the potential well. It determines the limiting momentum
$\rho_f\propto\sqrt{\varepsilon'_f}$.

Turning to a nuclear Fermi liquid (see Fig.~2b), we encounter the, strictly speaking,
inadequate nature of the concept of the ``bottom" of the potential well. Actually, a situation
with noninteracting quasiparticles can exist only in the immediate vicinity of the Fermi
limit. Even if we have in some sense complete data on the energy spectrum of the nucleus
(we assume that the energy $E\{n_{\bf k}\}$ of the body is known as a functional of the
occupation numbers $n_{\bf k}$), it is not possible to give a reasonable definition of the
quasiparticle energy $\varepsilon'$ relative to the ``bottom." Nevertheless, the idea of a
limiting momentum still makes sense. In principle we could deduce the value of $\rho_f$,
say, from the form of the wave function of the last quasiparticle.

However, if the ``bottom," strictly speaking, does not exist, and the quasiparticle energy
$\varepsilon'$ calculated with respect to it is not an adequate quantitative notion, then
what physically determines the limiting momentum $\rho_f$? The answer to this question is
given by the major assumption on which the theory developed in the next section is based.

\section{Theory of the shell structure of a spherical nucleus}

As a starting point for what follows we will assume that the value of $\rho_f$ is completely
determined by the total number of particles $N$. It is assumed that the function
$N(\rho_f)$ does not undergo oscillations\footnote{For the model of an ideal gas located in
an externally applied potential well, the
oscillations would be given by Eq.~(16).}. In a purely formal sense, only
differentiability of this function is required (see below, the transition from Eq.~(22) to
Eq.~(24)). However, in connection with criterion (2), it is natural to be interested in its
asymptotic representation for large $\rho_f$. In that case the corresponding expansion should
be terminated with a finite number of terms:
\begin{equation}\label{3}
    N(\rho_f)=\frac4{9\pi}\rho_f^3-s\rho_f^2+q\rho_f.
\end{equation}
Actually the fifth term of the expansion, which is $\sim\rho_f^{-1}$, would give physically
meaningless fractional additions to the number of particles. In a systematically macroscopic
treatment it is obviously required to discard also the fourth term as describing the
single-particle effect (see the discussion of this point in the Introduction). In what
follows we will also drop all similar non-macroscopic terms of the type
$\rho_f^0\sim1$\footnote{The correctness of the third term of Eq. (3) also could turn out
to be questionable,
since $2l + 1 \sim\rho_f$ particles fill a single level in a spherically symmetric field.
However, the unavoidable zero-point oscillations of the deformation have a scale $\Delta\alpha
\sim\rho_f^{-2}$, and the corresponding shift in particle energy is of the same order
of magnitude. The
degeneracy is as a result lifted to a sufficient degree, while, on the other hand, such
small deformations do not yet violate the conservation of the integral of motion $l$
(see ref.~[3]). Thus, the averaging of the function $N(\rho_f)$ implied in (3) is achieved
automatically in a real situation.}.

The assumptions described above have a great similarity to the system of assumptions of the
theory of an unbounded Fermi liquid [10,11].  In particular, the coefficient in the first
term of the right-hand side of (3) corresponds (if the additional spin doubling is taken into
account) to the volume contribution to the number of cells of phase space, that is, it is
equal to the corresponding expression for an ideal Fermi gas. However, we have no basis for
assuming that the remaining terms are also independent, to the same degree, of the law of
interaction between particles. On the contrary, the surface term $-s\rho_f^2$, and also
the curvature term $q\rho_f$, take into account the structure of the transition layer on
the nuclear surface, spin-orbit coupling, and so forth. Numerical values of the coefficients
$s$ and $q$ must be determined experimentally (see the following section).

We represent the ground-state energy of a spherical nucleus in the form of the sum
\begin{equation}\label{4}
    E=E_0+E_1
\end{equation}
of a smooth part $E_0$ and an oscillating part $E_1$. In the calculation of the latter,
it is convenient to express it initially in terms of the variable $\rho_f$.  The coupling
noted in the Introduction between the oscillations and quasi-particles approaching the Fermi
limit permits the problem to be reduced to the model [3] of a gas in a well with a constant
potential inside it (the last fact corresponds physically to uniformity of the spatial
distribution of matter in the internal region of the nucleus).  Subsequent use of Eq.~(1)
to carry out the summation over the quantum numbers $l$ and $n$ will lead to integrals of
two substantially different types. Those of them which are due to practically the entire
region $0 < \rho = kR < \rho_f$ of values of the particle wave number do not permit
generalization to a real Fermi liquid. However, they depend smoothly on $\rho_f$ and are not of
interest here---the problem of explicit calculation of the function $E_0(\rho_f)$ does not
arise here. The integrals which have an oscillating dependence on $\rho_f$ converge rapidly
at $\rho\simeq \rho_f$, and the generalization of their contribution to the case of a Fermi
liquid is obvious.  This fact even allows us to limit ourselves to calculation of the number
of particles
\begin{equation}\label{5}
    \widetilde{N}=\widetilde{N}_0+\widetilde{N}_1
\end{equation}
in this model. The point is that, in the terms which determine $\widetilde{N}_1$, the energy
of a single particle (subsequently a quasi-particle) would anyway be taken outside the integral
sign as a slowly varying function. Equation (5) corresponds to the quantity
\begin{equation}\label{6}
    \varphi(l,n)=(2l+1)w_f(l,n),
\end{equation}
which must be summed over $l$ and $n$. The Fermi distribution
\begin{equation}\label{7}
    w_f(l,n)=\Bigg\{
    \begin{array}{l}
    1\quad\mathrm{for}\quad \rho_{ln}<\rho_f,\\
    0\quad\mathrm{for}\quad \rho_{ln}>\rho_f,
    \end{array}
\end{equation}
depends on these same quantum numbers\footnote{In view of the macroscopic nature of the
effect being studied, the existence of spin in
the nucleon will be taken into account easily by subsequent doubling of the result. We note
that in problems of this type it is more natural to begin the enumeration of the principal
quantum number $n$ not from unit but from zero. For example, the $1s$ state is assigned to
$n = 0$. In particular, this provides the possibility of formally symmetric
application of Eq.~(1) to summation over the two quantum numbers (see Eq. (11) below).}
through the eigenvalues $\rho = \rho_{ln}$. The Bohr-Sommerfeld quantization rule [2] serves
to determine the eigenvalues,
\begin{equation}\label{8}
    \int_a^Rk_l(r)dr=\pi(n+\gamma),
\end{equation}
where the internal turning point $r = a$ is due to the centrifugal barrier, and $\gamma < 1$
defines an additional phase shift which depends on the nature of the boundary conditions. It
is not necessary to calculate integral (8) from
the beginning, since in this case both the wave functions themselves (spherical Bessel
functions) and their quasi-classical asymptotic forms are well known. Using a notation close
to that of ref.~[3], we have
\begin{equation}\label{9}
    \begin{split}
    \rho(\sin\beta-\beta\cos\beta)=\pi(n+3/4),\\
    \beta=\arccos\frac{l+1/2}{\rho}=\arcsin\sqrt{1-\frac{(l+1/2)^2}{\rho^2}}.
    \end{split}
\end{equation}
The Jacobian of the transformation to the new variables also was determined in ref.~[3]:
\begin{equation}\label{10}
    dndl=\frac1{\pi}\rho d\rho\sin^2\beta d\beta.
\end{equation}
Two-fold summation of the function (6) according to Eq.~(1) gives
\begin{equation}\label{11}
    \begin{split}
    \sum_{n=0}^{\infty}\sum_{l=0}^{\infty}\varphi(l,n)=\tfrac14\varphi(0,0)
    +\iint_0^\infty\varphi(l,n)dldn\\
    +\sum_{\nu=1}^{\infty}\iint_0^{\infty}\left(e^{i2\pi\nu n}+e^{-i2\pi\nu n}\right)\varphi(l,n)\,dldn
        +\sum_{\lambda=1}^{\infty}\iint_0^{\infty}\left(e^{i2\pi\lambda l}+e^{-i2\pi\lambda l}
    \right)\varphi(l,n)dldn\\
        +\frac12\left\{\int_0^{\infty}\varphi(0,n)dn+\sum_{\nu=1}^{\infty}\int_0^{\infty}
    \left(e^{i2\pi\nu n}+e^{-i2\pi\nu n}\right)\varphi(0,n)dn\right\}\\
        +\frac12\left\{\int_0^{\infty}\varphi(l,0)dl+\sum_{\lambda=1}^{\infty}\int_0^{\infty}
    \left(e^{i2\pi\lambda l}+e^{-i2\pi\lambda l}\right)\varphi(l,0)dl\right\}\\
    +\sum_{\nu=1}^{\infty}\sum_{\lambda=1}^{\infty}
    \int\!\!\!\int_0^{\infty}\left[e^{i2\pi(\lambda l+\nu n)}+e^{-i2\pi(\lambda l+\nu n)}
    +e^{i2\pi(\lambda l-\nu n)}+e^{-i2\pi(\lambda l-\nu n)}\right]\varphi(l,n)dldn.
    \end{split}
\end{equation}

All terms in the right-hand side of (11) permit analysis and classification on the basis of the
criteria mentioned above. The non-macroscopic nature of the first of these terms is obvious.
The non-oscillatory behavior of the second term follows directly from its form.
After carrying out the integration, we can verify that the third term also provides no
macroscopic contribution. As the result of integration on the basis of the same formulas (9)
and (10) and subsequent summation over $\lambda$, we find that the fourth term is $-\rho_f/6\pi$.
It depends smoothly on the limiting momentum.  The same can be said of the two pairs of
following terms, which are enclosed in curly brackets. In both cases the sum which adds to
the integral actually describes its increase as the result of those levels which, with further
advance of the Fermi limit, appear separately for $\rho<\rho_f$ \footnote{This rather formal
fact has already been mentioned in the Introduction and commented on in footnote $^5$.}.

The oscillations are described by the double sum over $\lambda$ and $\nu$ (the last term in
Eq.~(11)).  Before we begin calculation of this sum, we will make one remark of a less formal
nature. In addition to the Fermi limit $\rho=\rho_f$ (see Eq.~(7)) the distribution of particles
along the $l$ (or $\beta$) axis is also cut off---there are no negative orbital angular momenta.
An intuitive interpretation of this quantum phenomenon must lie in the fact that the shell
oscillations depend on the relative position of these two limits. As will be seen from Eq.~(19)
below, a magic nucleus arises when the two limits coincide. In
view of this role of the lower limit of the angular momentum scale, where $\beta\approx\pi/2$,
it is convenient to introduce an additional angle
\begin{equation}\label{12}
    \widetilde{\beta}=\pi/2-\beta.
\end{equation}
Then it is easy to see that the integral under the summation sign, generally speaking, has
the same structure as in the second term of the right-hand part of Eq.~(11), i.e., it is
non-macroscopic. However, a saddle point arises for a definite ratio between $\lambda$ and $\nu$
at the lower limit of the angular momentum axis, and this provides a macroscopic contribution
\footnote{This pertains to the first two exponential terms under the integral sign. Some other terms
of the double sum have saddle points at intermediate angular-momentum values, which do not
affect the result. It is interesting to note that in the preceding work devoted to calculation
of the rigidity of the spherical configuration of the simplest model [3] these same points
played the role of poles of the integrand, and incidentally gave likewise contribution.}.

If we take into account Eqs.~(9) and (12), we can write the expansion of the argument of the
exponential in powers of $\widetilde{\beta}$ up to and including quadratic terms as follows:
\begin{equation}\label{13}
    2\pi(\lambda l+\nu n)\approx-\pi\lambda-\tfrac32\pi\nu+2\nu\rho+\pi(2\lambda-\nu)
    \rho\widetilde{\beta}+\nu\rho\widetilde{\beta}^2.
\end{equation}
Vanishing of the linear term (the condition for existence of a saddle point at $\widetilde{\beta}=0$)
requires that
\begin{equation}\label{14}
    \nu=2\lambda.
\end{equation}
Since the phase-space element is given by Eq.~(10), we obtain from Eqs.~(6), (7), (9), and (12)
\begin{equation}\label{15}
    \iint e^{i2\pi(\lambda l+\nu n)}\varphi(l,n)\,dldn\approx\frac2{\pi}\int^{\rho_f}d\rho
    \rho^2e^{i4\lambda\rho}\int_0e^{i2\lambda\rho\tilde{\beta}^2}\widetilde{\beta}\,
    d\widetilde{\beta}=\frac{\rho_f}{8\pi}\frac{e^{i4\lambda\rho_f}}{\lambda^2}.
\end{equation}
Addition of the complex conjugate of the expression and summation over the single remaining
free index leads to
\begin{equation}\label{16}
    \widetilde{N}_1=\frac{\rho_f}{4\pi}\sum_{\nu=1}^{\infty}\frac{\cos4\nu\rho_f}{\nu^2}
\end{equation}
according to Eqs.~(5) and (6).  For conversion to the oscillating part of the energy of the
nucleus $E_1$ it is necessary to take into account the nucleon spin by doubling, and also to
multiply Eq.~(16) by the energy $\varepsilon_f=-\varepsilon$ ($\varepsilon$ is the nucleon
binding energy) of a single quasiparticle:
\begin{equation}\label{17}
    E_1=-\varepsilon\frac{\rho_f}{2\pi}{\EuFrak{ M}}(\rho_f).
\end{equation}

Thus, for an absolutely sharp step-shaped Fermi limit for the quasiparticles, the shell effects
in a spherical nucleus are described by a certain universal periodic function
\begin{equation}\label{18}
    \EuFrak{M}(\rho_f)=\sum_{\nu=1}^{\infty}\frac{\cos4\nu\rho_f}{\nu^2}
\end{equation}
whose plot is shown in Fig.~3. Its derivative has discontinuities at values of the argument
\begin{equation}\label{19}
    k_fR=\frac{\pi}2p,\qquad p=2,3,4,\ldots,
\end{equation}
which correspond to the magic cusps (see Eq.~(17)). In other words, Eq.~(19) is a unique
quantization rule for magic values of the parameter $\rho_f = k_fR$.

We emphasize that the theory being developed is an
expansion in inverse powers of the integer ``number" of the magic nucleus $p$ (see also Eq.~(2)).
It cannot pretend to give a quantitative description of the case of extremely small values,
especially since it is then evidently impossible in general to make any reasonable separation of
``smooth" and ``oscillating" effects. However, the number of nodes of the radial wave
function of the nucleon is such a stable characteristic that we can hope to trace it, at least
qualitatively, to the lightest nuclei.  Since $\rho_f = \pi$ already corresponds to a $1s$
state (the doubly magic nucleus $_2\mathrm{He}_2^4$ corresponds to this physically), the region of
possible values of $p$ should begin  with 2 \footnote{The definition of the effective radius $R$
of the nucleus implied in (19) concerns the
internal structure of the nucleus exclusively. Since we have reduced the problem to a model
with an impenetrable wall, we must represent the wall as constructed where the wave function
of the corresponding quasiparticle extrapolated from the internal region approaches zero.
In other words, this effective boundary of the nucleus can always be disposed so that the
additional phase shift from the Bohr-Sommerfeld rule (8), in accordance with (9), will take
on the value $\gamma = 3/4$ for the quasiparticles playing the principal role. In close
connection with this fact, the equations expressing the shell oscillations in terms of $\rho_f$
are universal; they do not depend on the spin-orbit interaction or on the structure of the
surface layer where. After transformation to the $N$-scale this universality
is lost (see also Eq.~(3) and the explanation for it).}.

It is easy to verify that the right-hand part of Eq.~(18) is a Fourier series for an elementary
function which we will write out explicitly in a form valid for the two periods adjacent to
the magic number $p$:
\begin{equation}\label{20}
    \EuFrak{M}(\rho_f)=\frac{\pi^2}6-2\pi\left|\rho_f-\frac\pi2p\right|+4\left(\rho_f-\frac\pi2p
    \right)^2,\quad \frac\pi2(p-1)<\rho_f<\frac\pi2(p+1).
\end{equation}

The absolute-value sign reflects the non-analyticity of the function at the magic cusp. Using
the indices $+$ and $-$ to distinguish the values of the discontinuous function to the right
and left respectively, we have
\begin{equation}\label{21}
    \left.\frac{d\EuFrak{M}}{d\rho_f}\right|_{\pm}=\mp2\pi.
\end{equation}
Now, returning to (17), it is not hard to obtain with the required accuracy an expression for
the discontinuity in the derivative of the oscillating part of the energy of the nucleus:
\begin{equation}\label{22}
    \Delta\left(\frac{dE_1}{d\rho_f}\right)=\rho_f(\varepsilon_++\varepsilon_-)=2\rho_f\bar{\varepsilon}.
\end{equation}

In order to convert from the limiting momentum to the true number of particles (3), it is
necessary to multiply both sides by $d\rho_f/dN$. At the same time we recognize that the smooth
component $E_0$ does not have a singularity at the cusp, so that Eq.~(22) actually gives the
discontinuity in the derivative of the entire energy $E$:
\begin{equation}\label{23}
    \frac{d\rho_f}{dN}\,\Delta\left(\frac{dE}{d\rho_f}\right)=2\rho_f\frac{d\rho_f}{dN}\bar{\varepsilon}.
\end{equation}
This relation for the discontinuity in the chemical potential can be given a more compact form:
\begin{equation}\label{24}
    \Delta\left(\frac{dE}{dN}\right)=\frac{\bar\varepsilon}{dN/d(k_fR)^2}.
\end{equation}
This can also be considered as a formula for the discontinuity in nucleon binding energy
\begin{equation}\label{25}
    \varepsilon=-\frac{dE}{dN}
\end{equation}
in the vicinity of a magic nucleus.  Then, obviously, we must assume $\Delta\varepsilon\equiv
\Delta\left({dE}/{dN}\right)=\varepsilon_--\varepsilon_+$.

\section{Comparison with experiment}

The most direct physical meaning is attached to the number-of-particles scale, as the most
closely connected with experiment. We will discuss this after we have determined for the
nuclear Fermi liquid, from experimental data, the numerical values of the coefficients $s$
and $q$ entering into Eq.~(3). If we assume that, of the magic numbers found experimentally
up to the present time, the most important in this connection are 82 and 126, then according
to (19) and (3) these are sufficient to determine the two parameters. We finally obtain
\begin{equation}\label{26}
    s=1.1,\qquad q=6.8.
\end{equation}
It is interesting to note that the surface term exceeds by only a factor of about two the
value $\tilde{s} = 1/2$ characteristic of an ideal Fermi gas adjacent to an impenetrable
wall [12]. With the values in (26), Eqs.~(3) and (19) for the number of particles
in a magic nucleus take the form
\begin{equation}
N = 0.548p^3 - 2.79p^2 + 10.7p.
\end{equation}
The results of a calculation on the basis of Eq.~(27) are compared with the known magic numbers
in Table~I. Down to $N = 28$ inclusive, the agreement must be considered satisfactory.  For the
lightest magic nuclei there is not even qualitative agreement
\footnote{The frequently encountered separation of a single  $1f_{7/2}$ level in an individual
shell has always been problematical. The reasoning and formulas of the preceding section
create the impression of rejecting the magic number 20 which arises here.
It is necessary, however, to make the reservation that magic phenomenon can in any case
not be regarded as macroscopic when it comes to the lightest nuclei.}.

If we take into account the two-component nature of nuclear matter, the oscillating term is
given by the sum of expressions (17) for neutron and proton quasiparticles
\begin{equation}\label{28}
    E_1(N,Z)=-\varepsilon_N\frac{\rho_f^N}{2\pi}{\EuFrak{ M}}(\rho_f^N)
    -\varepsilon_Z\frac{\rho_f^Z}{2\pi}{\EuFrak{ M}}(\rho_f^Z).
\end{equation}

%\begin{minipage}
\begin{table}
\begin{tabular}{|c|c|c|}
\hline
\multicolumn{3}{|c|}{\bf Table I}\\
\hline
$p$ & $N$ & $N_{theor}$\\
\hline
2 &  2 &  15 \\
3 &  8 &  22 \\
4 &  28 & 33 \\
5 &  50 & 52 \\
6 &  82 & 82 \\
7 &  126 & 126 \\
8 &  184(?) & 187 \\
9 &    &  270 \\
10 &   &  376 \\
\hline
\end{tabular}
\end{table}
%\end{minipage}

We note in this connection that the agreement observed so far of the corresponding magic
numbers evidently indicates practically identical values of the parameters $s$ and $q$ for
the two components. We may suppose that this is due to the relative smallness of the effects
distinguishing protons and neutrons in the nucleus, such as, for example, the Coulomb curve
(the radial dependence of electrostatic potential in the internal region of the nucleus)
\footnote{These remarks have pertained to the total number (3) of nucleons of a given kind in the
nucleus. However, in regard to the more detailed properties of the residual interaction
of quasiparticles close to the Fermi limit, the neutron and proton components may be far
from identical; see below.}.

Clarification of the nature of the singularity in the energy $E(N, Z)$ at the point of
occurrence of a doubly magic nucleus presents substantial theoretical interest. If we return
to expression (20) and neglect in it the last quadratic term, we see that, according to (28),
the surface $E_1(N, Z)$ has here the form of a pyramid of rhombic cross section with its axis
directed vertically downward. As far as we can judge from the experimental data, the shape of
the intersection of the mass surface with the plane $E_1 = \mathrm{const}$ is actually close
to a rhombus near a doubly magic nucleus; see, for example, ref.~[13].

Incidentally, the separation of the oscillating part $E_1$ of the mass, from a purely
practical point of view, suffers from a certain ambiguity. Therefore our attention is drawn
to the relation (24), which does not depend on this procedure, for the discontinuity in
binding energy of a nucleon of the corresponding type in a singly magic nucleus.
Comparison with mass data (see, for example, ref.~[14]) has shown that the theory based on
the step-shaped distribution (7) does not agree with experiment for the magic numbers found
up to the present time. Since the oscillations are due to the sharpness of the Fermi
distribution, it is necessary to have in mind that the jump in quasiparticle occupation
numbers from unity to zero is the maximum permissible by the Pauli principle.
Consequently the oscillations calculated according to Eq.~(7) are exaggerated, and the
value $(\Delta\varepsilon)_0$ calculated from Eq.~(24) actually exceeds everywhere the really
observed binding-energy jump $\Delta\varepsilon$. It is natural to characterize this discrepancy
by the quantity
\begin{equation}\label{29}
    \omega=\frac{(\Delta\varepsilon)_0}{\Delta\varepsilon}-1,
\end{equation}
which would approach zero in the ideal case (7) of absence of residual interaction between
quasiparticles. Values of $\omega$ for 52 magic nuclei are listed in Table~II
\footnote{The well known even-odd oscillations of nuclear masses are not related to the phenomenon
being discussed. However, the widely used methods of calculating the corresponding
correction (see, for example, ref.~[15]) are purely empirical and can hardly be considered
reliable. We therefore selected only those cases in which there are data on the binding
energy of two nucleons. On removal of two nucleons of the same kind, the type of nucleus
does not change, and the even-odd correction to the mass drops out of the result.}.

\begin{center}
\begin{table}
\begin{tabular}{|c|c|c|c|c|c|c|c|}
\hline
\multicolumn{8}{|c|}{\bf Table II}\\
\hline
\multicolumn{8}{|c|}{Neutron magic nuclei} \\
\hline
\multicolumn{2}{|c|}{$p=4$; $N=28$} &
\multicolumn{2}{|c|}{$p=5$; $N=50$} &
\multicolumn{2}{|c|}{$p=6$; $N=82$} &
\multicolumn{2}{|c|}{$p=7$; $N=126$} \\
\hline
$Z$  &  $\omega$ &
$Z$  &  $\omega$ &
$Z$  &  $\omega$ &
$Z$  &  $\omega$ \\
\hline
22 & 3.64 &  36 & 2.64  &  54 & 1.25 &  81 & 0.65 \\
23 & 4.76 &  37 & 2.57  &  55 & 1.18 &  82 & 0.56 \\
24 & 6.20 &  38 & 2.28  &  56 & 1.38 &  83 & 0.72 \\
25 & 7.26 &  39 & 2.15  &  57 & 1.66 &  84 & 1.06 \\
26 & 7.44 &  40 & 2.62  &  58 & 2.06 &  85 & 1.14 \\
   &      &  41 & 2.76  &  59 & 2.18 &  86 & 1.72 \\
   &      &  42 & 3.02  &  60 & 2.42 &  87 & 1.96 \\
   &      &     &       &  61 & 2.49 &  88 & 1.95 \\
\hline
\multicolumn{8}{|c|}{Proton magic nuclei}\\
\hline
\multicolumn{2}{|c|}{$p=4$; $Z=28$} &
\multicolumn{2}{|c|}{$p=5$; $Z=50$} &
\multicolumn{2}{|c|}{$p=6$; $Z=82$} &
\multicolumn{2}{|c|}{} \\
\hline
$N$  &  $\omega$ &
$N$  &  $\omega$ &
$N$  &  $\omega$ & & \\
\hline
31 & 2.10 &  64 & 1.42 &  116 & 0.52 & &\\
32 & 2.32 &  65 & 1.20 &  117 & 0.37 & &\\
33 & 2.68 &  66 & 1.59 &  118 & 0.73 & &\\
34 & 2.66 &  67 & 1.67 &  119 & 0.83 & &\\
35 & 3.01 &  68 & 1.99 &  120 & 0.63 & &\\
   &      &  69 & 1.98 &  121 & 0.65 & &\\
   &      &  70 & 2.23 &  122 & 0.81 & &\\
   &      &  71 & 2.50 &  123 & 0.68 & &\\
   &      &     &      &  124 & 0.65 & &\\
   &      &     &      &  125 & 0.51 & &\\
   &      &     &      &  126 & 0.49 & &\\
\hline
\end{tabular}
\end{table}
\end{center}

Each magic number is associated with several data differing in the number of nucleons of
different kinds. Although some systematic behavior is apparently observed in the dependence
on the latter, we will assume tentatively in each case that these are deviations of an
accidental and uncorrelated nature.  The corresponding average values of $\omega$ are shown
in Fig.~4, together with the corresponding variances. The
rapid decrease of $\omega$ in the transition to heavy magic nuclei reflects the
naturally expected drop in strength of the residual interaction with increasing nuclear size.
The electrically charged protons are less affected by the residual interaction from the very
start, and at the end of the periodic table it shows a sharper drop than in the case of neutrons.
We will dwell briefly on the attempts to provide some simple empirical description of the
behavior of the points plotted in Fig.~4.  From the number of curves that asymptotically
approach $\omega = 0$, the possibility $\omega = C/p^3$ is suggested. For neutrons this
gives $C = 336 \pm12$; according to the criterion of likelihood [16] of the law
$\omega\propto 1/p^3$, the probability of $\chi^2$ greater than the value observed in the
present case is 20\%. However, in the case of protons the corresponding probability of a
similar law drops to 0.5\%. Another possibility is that the proton curve $\omega(p)$
intersects the axis of abscissas, and we will discuss this in the next section.

\section{Discussion}

The problem of the energy spectrum of a quantum liquid consisting of fermions is rather
complex in itself. In application to such a peculiar object as a nucleus it becomes
particularly difficult and many-faceted. The accuracy of the description of its excitations
in the model of an ideal gas of quasiparticles is limited by the residual interaction
between them. It is natural to suppose that it is created only by the finite size of the
system, and that in the transition to unbounded nuclear matter we would obtain the simplest
variety of Fermi-liquid spectrum [10,11].  Another variant of the energy spectrum takes
into account the possibility that, even in an infinite quantum liquid, residual phenomena
such as the Cooper phenomenon [11] are preserved. We must suppose that the deviation shown
in Fig.~4 from the asymptotic value $\omega= 0$ is not accidental. In spherical nuclei,
which owe their very existence to the residual
interaction, the latter must of course smear in some way the boundary of the Fermi distribution,
and in so doing suppress the oscillations. As a result of this special role and, apparently,
of the relatively large magnitude of the effect, it will probably be possible to clear up
the question of the residual interaction, which is finally and with difficulty yielding to analysis.

The monotonic drop observed so far (Fig.~4) in the function $\omega(p)$ already permits some
preliminary conclusions to be drawn. The point is that the shell oscillations calculated in
the present article are an extremely fine instrument for analysis of the quasiparticle
distribution near the limit $\rho\approx\rho_f$, and with increasing nuclear size the accuracy
of this method is increased. In fact, according to (1), (9), (11), and (15), the oscillations
penetrate below the Fermi limit by a depth $\Delta\rho\sim 1$, i.e., the relative thickness of
the layer of the Fermi distribution occupied by them is $\sim1/\rho_f\propto1/R$. On the other
hand, in the Cooper phenomenon the width of the zone of smearing of the distribution of
initial quasiparticles does not depend on the size of the system. It can be seen from this
how greatly the oscillations would be weakened in the presence of such a phenomenon. The drop
in the $\omega(p)$ curve would be delayed and, furthermore, it evidently would have the reverse
behavior. If still heavier magic nuclei do not exhibit similar features and the monotonic nature
of the $\omega(p)$ function is further confirmed, it will be necessary to give a negative answer
to the question of the Cooper phenomenon (or any other similar instability of an absolutely
sharp Fermi limit of quasiparticles in infinite nuclear matter).

In speaking of the residual interaction, it is necessary to recall that not all conceivable
varieties of this interaction are, so to speak, physically real. At least some variants of
the interaction may be excluded by a canonical transformation to new quasiparticles without
loss of sharpness in the Fermi limit of their statistical distribution. The well-known case
of repulsive forces of zero radius between fermions [11] serves as an example of this.
This tendency of a repulsive interaction to be ``renormalized" may turn out to be important
in regard to the proton component of nuclear matter. Actually, the points referring to protons
in Fig.~4 may reflect a competition between the nuclear and Coulomb interactions. If the
latter become dominant with increasing nuclear size, this may be expressed in the intersection
of the corresponding $\omega(p)$ curve with the abscissa.  This intersection would be a
termination point for the function $\omega(p)$, beyond which the quasiparticles would satisfy
the ordinary Fermi distribution (7). On the other hand, according to the results of ref.~[3],
disappearance of the residual interaction between proton quasiparticles would not, apparently,
favor the stable equilibrium nature of a spherical configuration. Even in the already studied
region of nuclei beyond the doubly-magic lead, the data on the phase transition [4] appear rather
symptomatic in this respect. Although the entire period from 82 to 126
contains 44 protons, addition to lead of only five of them turns out to be sufficient to
remove the stability of the spherical nuclear shape. We may suppose that beyond the terminal
point of the proton $\omega(p)$ curve the regions of existence of spherical nuclei narrow more
rapidly or even will not exist at all.

We express our thanks to M.Ya. Amusya, A.I. Baz', B.L. Birbrair, I.I. Gurevich, M.V. Kazarnovskii,
L.P. Kudrin, I.M. Pavlichenkov, G.A. Pik-Pichak, V.P. Smilga, V.V. Sudakov, and O.B. Firsov for
discussion of the results of this work. We also express our indebtedness to I.M. Lifshitz for
pointing out to one of the authors (V.N.) the mathematical apparatus which apparently turns out
to be most convenient for discussion of problems of this nature.

\vspace{3cm}

\centerline{\bf Figures captions}

\bigskip
\bigskip

Fig.~1. The smooth part of the mass, described by the Weizs\"acker formula, does not have
singularities; it is taken as the reference zero of the ``shell" effect pictured here.
A more detailed and accurate plot would contain a whole set of curves corresponding to the
different chemical elements (see, for example, the review of Myers and Swiatecki [13]).
As a function of the number of protons $Z$ the behavior of the mass reveals no less sharply
expressed features of a similar nature; see also Sections 3 and 4 of the present article.

\bigskip

Fig.~2.

\bigskip

Fig.~3.

\bigskip

Fig.~4. Values of the quantity $\omega$: $\circ$ --- neutron magic numbers, $\bullet$ --- proton
magic numbers.

\end{document}